# Citation importance-aware document representation learning for large-scale science mapping

Zhentao Liang[a,b], Nees Jan van Eck[c], Xuehua Wu[a,b], Jin Mao[a,b,*], Gang Li[a,b]

[a] Center for Studies of Information Resources, Wuhan University, Bayi Rd 299, Wuhan 430072, China
[b] School of Information Management, Wuhan University, Bayi Rd 299, Wuhan 430072, China
[c] Centre for Science and Technology Studies, Leiden University, Kolffpad 1, Leiden 2332BN, Netherlands
* Corresponding author. Email: maojin@whu.edu.cn

**Abstract:**
Effective science mapping relies on high-quality representations of scientific documents. As an important task in scientometrics and information studies, science mapping is often challenged by the complex and heterogeneous nature of citations. While previous studies have attempted to improve document representations by integrating citation and semantic information, the heterogeneity of citations is often overlooked. To address this problem, this study proposes a citation importance-aware contrastive learning framework that refines the supervisory signal. We first develop a scalable measurement of citation importance based on location, frequency, and self-citation characteristics. Citation importance is then integrated into the contrastive learning process through an importance-aware sampling strategy, which selects low-importance citations as "hard negatives". This forces the model to learn finer-grained representations that distinguish between important and perfunctory citations. To validate the effectiveness of the proposed framework, we fine-tune a SciBERT model and perform extensive evaluations on SciDocs and PubMed benchmark datasets. Results show consistent improvements in both document representation quality and science mapping accuracy. Furthermore, we apply the trained model to over 33 million documents from Web of Science. The resulting map of science accurately visualizes the global and local intellectual structure of science and reveals interdisciplinary research fronts. By operationalizing citation heterogeneity into a scalable computational framework, this study demonstrates how differentiating citations by their importance can be effectively leveraged to improve document representation and science mapping.

# 1 Introduction

Scientific innovation is a major driving force for economic growth and social development. According to Schumpeter (1934), the new combination of existing production factors is key to innovation. This theory has also been extended to scientific research, where novel discoveries primarily build on the synthesis of existing knowledge (Lin et al., 2025; Wang et al., 2017; Uzzi et al, 2013). However, effective synthesis of scientific knowledge requires identifying relevant prior work, a task that has become increasingly challenging due to the exponential growth of scientific literature in recent decades (Bornmann et al., 2021).

To address this challenge, science mapping has emerged as an important approach to understanding the research landscape (Wang et al., 2023; Liu et al., 2023). A map of science reveals the disciplinary or thematic structure of science (Waltman & van Eck, 2012; van Eck & Waltman, 2017), helping scientists navigate the vast universe of scientific literature. It also serves as the foundation for downstream tasks such as research front identification (Liang et al., 2021; Huang et al., 2021), topic evolution analysis (Hu et al., 2019; Huang et al., 2024), and academic recommendation (Liu et al., 2023; Zhao et al., 2016). The effectiveness of these tasks, however, depends on the accuracy of science mapping, which in turn is determined by the quality of scientific document representations and inter-document relationships.

Traditionally, two main approaches have been widely adopted for perform science mapping: citation-based and text-based methods (Bascur et al., 2023; Chandrasekharan et al., 2021; Boyack & Klavans, 2020; Janssens et al., 2006). Citations provide the most straightforward way to establish connections among scientific documents, as they are considered explicit indicators of intellectual acknowledgment and knowledge diffusion (Tahamtan & Bornmann, 2018). Co-citation and bibliographic coupling are introduced to complement direct citation relationships (Eto 2019; Bu et al., 2020; Kleminski et al., 2022). Nevertheless, some researchers have raised concerns about the complex motivations behind citations, which may undermine the accuracy of science mapping (Ding et al., 2014; Lyu et al., 2021). A recent survey (Teplitskiy et al., 2022) shows that many citations are made strategically and therefore the cited papers have little to no intellectual influence on the citing paper. In contrast, text-based approaches focus on the content of scientific documents, allowing them to identify potentially related works regardless of citation relationships (Waltman et al., 2020; Boyack & Klavans, 2020; Janssens et al., 2006). More recently, representation learning techniques have provided a promising way to capture the semantic information of scientific documents (Beltagy et al., 2019; Lee et al., 2020; Shen et al., 2023). However, disregarding citation information can harm the performance of document representation models. It is difficult to make fine-grained distinctions between semantically related documents as they are projected to a narrow area of the embedding space (Gao et al., 2019). Despite some limitations, citations remain a valuable signal of relatedness. Therefore, integrating citation and semantic information is an important direction for improving the quality of document representations and the accuracy of science mapping.

Contrastive learning stands out as an effective approach in this direction (Gao et al., 2021). In a pioneering work, Cohan et al. (2020) proposed SPECTER, a transformer-based model that fine-tunes document representation through citation-informed contrastive learning. Documents connected by citations are treated as positive pairs and should be projected closer in the embedding space, whereas other document pairs should be pushed apart. While SPECTER is one of the state-of-the-art (SOTA) models for scientific document representation, the simple assumption that all citations are equal limits its ability to capture fine-grained document relatedness (Ostendorff et al., 2022). Perfunctory citations may introduce noise into the representations, as the model is forced to generate similar embeddings for weakly related documents. Although some researchers have attempted to differentiate citations by their function, sentiment, and importance (Aljuaid et al., 2021; Zhang et al., 2022; Budi & Yaniasih, 2023), these approaches remain experimental and face challenges in scaling to large applications. Therefore, effective integration of citation information into document representation learning, especially considering the heterogeneity of citations, remains a significant yet underexplored challenge in science mapping.

## 1.1 Research objective and contributions

This study aims to improve the accuracy of science mapping by enhancing the quality of scientific document representation. Our approach is driven by a central principle in scientometrics that not all citations are equal. Rather than attempting to achieve SOTA performance for all types of scientific document representation tasks, our objective is to investigate whether this scientometric principle can be operationalized to improve the performance of existing approaches that do not take citation heterogeneity into account. Our ultimate goal is to construct a more accurate map of science based on these improved representations. Driven by this application-oriented focus, this study prioritizes performance gains in tasks directly related to science mapping.

To achieve this objective, we propose an importance-aware fine-tuning approach that extends the original citation-informed contrastive learning framework (Cohan et al., 2020). Specifically, the representation learning model is designed not only to distinguish between documents with and without citations, but also to differentiate the degree of importance among citation pairs. Three key questions are addressed in this research:

**RQ1:** How can the importance of citations be quantified and integrated into the contrastive learning framework?

**RQ2:** Does the proposed importance-aware sampling strategy improve the quality of scientific document representation compared to baselines that treat all citations equally?

**RQ3:** To what extent does importance-aware scientific document representation improve the accuracy of science mapping?

To answer these questions, we first propose a citation importance measurement based on citation location, citation frequency, and self-citation characteristics. This information is then integrated into the contrastive learning process through an importance-aware sampling strategy, in which document pairs with higher citation importance scores are considered as positive. A dataset consisting of 4 million full-text articles and 67 million citations is used to construct training and development data. We fine-tune the SciBERT model with the proposed approach and evaluate the performance on SciDocs (Cohan et al., 2020) and the PubMed dataset (Ahlgren et al., 2020). Finally, the trained model is applied to a large-scale science mapping task covering 33 million papers across all disciplines from 2000 to 2022.

This study contributes to the field of information science in three ways:

(1) **A scalable framework for improving scientific document representation by**

**differentiating citation importance**. We propose a novel importance-aware sampling strategy to refine the supervisory signals in contrastive learning. Unlike previous approaches that treat citation relationships equally, our framework explicitly captures citation heterogeneity using scalable, metadata-based features. This allows the model to learn finer-grained representations by distinguishing important citations from perfunctory ones, offering an efficient solution for large-scale applications.

(2) **Extensive evaluation of the effectiveness and robustness of the proposed framework**. The model trained with our framework improves both scientific document representation quality and science mapping accuracy. Alternative implementations of our citation importance measurements and ablation studies confirm the robustness of these performance gains.

(3) **A large-scale, interdisciplinary application of the model trained with our framework**. We apply the trained model to construct global maps of science with different overlays, covering over 33 million scientific publications. The resulting maps accurately visualize the intellectual structure of science and reveal interdisciplinary research fronts.

These findings point to a promising direction for achieving a more accurate and comprehensive understanding of the structure of scientific knowledge and the global science landscape.

The remainder of this paper is organized as follows. Section 2 reviews related works. Section 3 presents the proposed methodology for enhancing the quality of scientific document representation and the accuracy of science mapping. Section 4 describes the experimental setup, and Section 5 reports the results. Section 6 discusses the theoretical and practical implications. Finally, Section 7 concludes the paper.

## 2 Related work

### 2.1 Mapping the structure of science based on scientific documents

Science mapping is a methodology used to analyze the structure of science through visual representations of relationships between scientific entities, such as papers, journals, disciplines, and scientists (Chen, 2017). These visualizations, often referred to as "science maps", provide insights into the organization of scientific knowledge, research trends, and collaborative patterns (Fortunato et al., 2018; Murray et al., 2023; Xie & Waltman, 2025). Among the various types of scientific entities, scientific documents are the primary carriers of research findings and the foundation from which other representations of the structure of science are derived (Chen, 2017). Therefore, depicting the structure of science through inter-document relationships and improving their accuracy has been a central focus of information science researchers (Waltman & van Eck, 2012; Ahlgren et al., 2020; Bascur et al., 2025).

Citations are the most established and widely used source for constructing inter-document relationships. According to normative citation theory, scholars cite prior works that are relevant or influential to their research (Merton, 1973; Tahamtan & Bornmann, 2018). Therefore, scientific documents can be positioned within a conceptual space based on citations, forming a citation network that represents the structure of science (Waltman & van Eck, 2012; van Eck & Waltman, 2017). Other citation-based methods have been proposed to complement direct citations, including co-citations, bibliographic coupling, and triangular citations (Eto 2019;

Kleminski et al., 2022; Liu et al., 2021). These approaches infer relationships between documents without direct citations, as well as the strength of such relationships, based on shared references or citing documents. However, actual citation practices are complex and often diverge from this idealized view (Tahamtan & Bornmann, 2019). Some documents are cited for strategic reasons, while relevant documents may be overlooked due to biases or oversight (Tahamtan et al., 2016; Kousha & Thelwall, 2024).

Text-based approaches offer a complementary way to establish inter-document relationships (Boyack & Klavans, 2020). Natural language processing techniques, such as lexical analysis, topic modeling, and representation learning, are used to identify semantic similarities between documents (Suominen & Toivanen, 2016; Hu et al., 2019; Long et al., 2025). By focusing directly on the content of scientific documents, these approaches can reveal thematic connections that are independent of citation links. However, text-based approaches often struggle with the nuances of scientific language (Li et al., 2023) and may fail to capture the intellectual lineage that citations more explicitly signify. While citations are noisy, reflecting diverse citing motivations (Tahamtan & Bornmann, 2019; Teplitskiy et al., 2022), they still function as a form of expert-annotated signal of relatedness. Therefore, hybrid approaches that integrate citation and semantic information represent a promising direction for improving the accuracy of science maps (Boyack & Klavans, 2020).

## 2.2 Representation learning of scientific documents

Representation learning aims to project scientific documents into a low-dimensional vector space, where the semantic or structural relationships between documents can be effectively captured and computed (Singh et al., 2023). Depending on the type of information used, document representation learning approaches can be broadly divided into three categories: semantic, structural, and hybrid approaches (Kozlowski et al., 2021).

Semantic approaches learn representations based on the textual content of scientific documents. Traditional methods, including TF-IDF, Latent Dirichlet Allocation (LDA), word2vec, and doc2vec, are primarily based on the bag-of-words assumption (Blei et al., 2003; Mikolov et al., 2013; Le & Mikolov et al., 2014). More recently, pre-trained language models (PLMs) based on the transformer architecture have been proposed to capture deeper semantic information of texts (Vaswani et al., 2017; Devlin et al., 2019). Built on the famous BERT model (Devlin et al., 2019), several variants have also been developed for scientific domains, such as SciBERT (Beltagy et al., 2019), BioBERT (Lee et al., 2020), and SsciBERT (Shen et al., 2023). While deemed powerful, the document representations generated by PLMs suffer from the issue of anisotropy. Specifically, the representations only occupy a narrow cone in the embedding space, which limits the ability to make fine-grained distinctions between similar scientific documents (Gao et al., 2019; Li et al., 2020).

Structural approaches primarily focus on the citation networks. Techniques such as matrix decomposition, random walk, and graph neural networks are employed to generate representations that preserve the structural proximity of documents (Goyal & Ferrara, 2018; Grover & Leskovec, 2016; He et al., 2023). Compared with semantic approaches, structural approaches are better at capturing implicit knowledge linkages that are not evident from textual similarity. However, this also implies that structural approaches inherit the limitations associated with citations (Kozlowski et al., 2021). For instance, the representation quality tends to be biased towards a few highly cited documents, and strategic and perfunctory citations introduce noise into the representation learning process (Ostendorff et al., 2022).

Hybrid approaches that integrate textual and citation information have been proposed to mitigate the abovementioned limitations. One common strategy is to incorporate content features as node attributes in graph-based models. For instance, textual features extracted by PLMs can be used to initialize node embeddings in graph neural networks (He et al., 2023; Xie et al., 2021; Zhu et al., 2021). Another promising direction, which is adopted by this study, is to integrate citation structures into the training process of PLMs. Contrastive learning has emerged as a powerful framework for this purpose (Gao et al., 2021). In this framework, documents connected by citations are treated as positive pairs, while others are treated as negative pairs. PLMs are subsequently optimized to minimize the distance between embeddings of positive pairs, while maximizing it for negative pairs, thereby mitigating the anisotropy problem (Cohan et al., 2020; 2023; Ostendorff et al., 2022). However, the heterogeneity of citations remains largely overlooked.

## 2.3 Differentiation of citations

Recognizing that not all citations are equal is essential for understanding relationships between scientific documents (Ding et al., 2014; Nicholson et al., 2021). Citations can serve different functions, ranging from providing background information and methodological foundation to comparing, criticizing, or simply mentioning previous work (Tahamtan & Bornmann, 2018; Zhang et al., 2022). They also vary in terms of sentiment and importance (Aljuaid et al., 2021; Budi & Yaniasih, 2023). To differentiate citations, researchers have focused on features extracted from full-text documents. For instance, Zhang et al. (2022) classify citation functions using deep learning models trained on document titles, sections, and citation context. Nazir et al. (2020) identify important citations by exploiting section-wise in-text citation frequencies and similarity measures. While these previous studies have successfully distinguished various facets of citations, it is challenging to scale these fine-grained analyses to large datasets due to computational complexity. Moreover, reliance on full-text data also limits applicability, since most scholarly datasets only contain metadata (Boyack et al., 2018).

Therefore, this study focuses on developing a scalable proxy for assessing citation importance that can be readily transformed into contrastive learning signals. By requiring PLM to distinguish between cited documents of varying importance, we aim to guide the model to generate finer-grained representations that, in turn, can be used to construct a more accurate map of science.

## 3 Methodology

To enhance the representation quality of scientific documents, we propose a citation importance-aware contrastive learning framework. The overall framework is illustrated in Fig. 1. Contrastive learning aims at optimizing the encoder to project similar documents closer in the embedding space while pushing dissimilar ones further apart. A crucial process in contrastive learning is defining similarity or relevance, which provides supervisory signals for model training. In the case of scientific documents, citation relationships offer a natural, albeit noisy, source of such signals (Cohan et al. 2020).

Our key innovation lies in moving beyond the simple assumption that all citations represent equal relatedness. Specifically, we introduce a method to assess citation importance and integrate it into the contrastive learning process through an importance-aware triplet sampling strategy. This forces the model to disentangle nuanced relationships, leading to improved document representations.

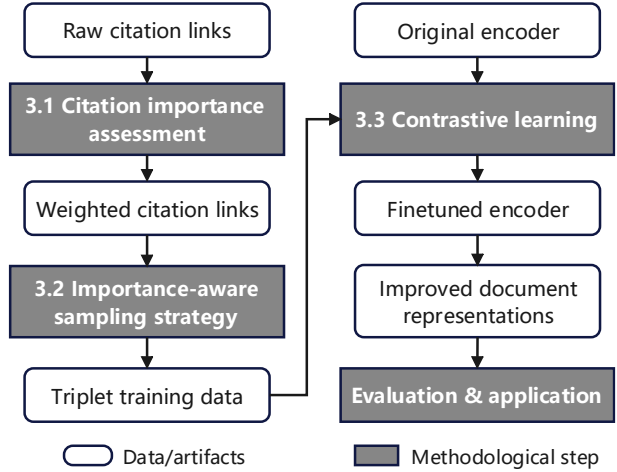

**Fig. 1.** Overview of the citation importance-aware contrastive learning framework.

## *3.1 Assessment of citation importance (RQ1)*

To answer the first research question, citation importance is calculated for each citing and cited document pair. Recent studies have identified several indicators of citation importance, including the cited location, frequency, sentiment, independence, and self-citation (Nazir et al., 2020; Aljuaid et al., 2021; Qavyyum et al., 2022). Among these indicators, location, frequency, and self-citation are considered most informative. These indicators are also suitable for large-scale applications because they do not involve in-depth analysis of citation contexts. Therefore, we propose a method to assess citation importance using the following equation:

$$importance_{i,j} = w_I f_I + w_R f_R + w_D f_D + w_S S \qquad (1)$$

For each citation pair ($i$, $j$), the importance score is a weighted sum based on how often document $j$ is cited in specific IMRaD sections of citing document $i$, plus an indicator for whether it is a self-citation. In equation (1), $f_I$, $f_R$, and $f_D$ denote the cited frequencies in the Introduction, Results, and Discussion & Conclusion sections, respectively. We exclude citations in the Method section, as they disturb our focus on the thematic relevance of scientific documents. To classify the IMRaD section types, we develop a SciBERT-BiLSTM-CRF model based on section title sequences. This composite neural network architecture is well-suited for sequence labeling tasks such as section classification, achieving a macro-F1 of 95.59% on our annotated dataset of 4000 papers (26,792 section titles).

$S$ is the self-citation indicator, which is equal to 1 if the citing and cited documents share at least one author, and 0 otherwise. For author disambiguation, we employ the author dataset processed by the Centre for Science and Technology Studies (CWTS) at Leiden University, which has a reported precision of 95% and is updated regularly (Caron & van Eck, 2014).

Each component is then multiplied by its corresponding weight ($w_I$, $w_R$, $w_D$, and $w_S$). These weights are estimated using the entropy weight method, which assigns weights objectively based on variability (Zhu et al., 2020).

*3.2 Importance-aware triplet sampling strategy*

To transform citation links and their importance into actionable supervisory signals, we propose an importance-aware sampling strategy for constructing training samples. Each training sample is a triplet ($p_a$, $p_p$, $p_n$), consisting of an anchor document $p_a$, a positive document $p_p$, and a negative document $p_n$. Anchor documents are randomly sampled from the dataset. Positive documents are selected from those cited by the anchor, but unlike previous research, which includes all cited documents as potential positive samples (as depicted in Fig. 2a), we select only those with the highest importance. Similarly, negative documents not only include uncited documents (easy negatives, $p_{en}$) but we introduce also "hard negatives" ($p_{hn}$): documents that are cited by the anchor but with the lowest importance. As illustrated in Fig. 2b, by treating these perfunctory citations as negatives, the model is forced to learn representations that can distinguish the nuanced differences between important and unimportant citations.

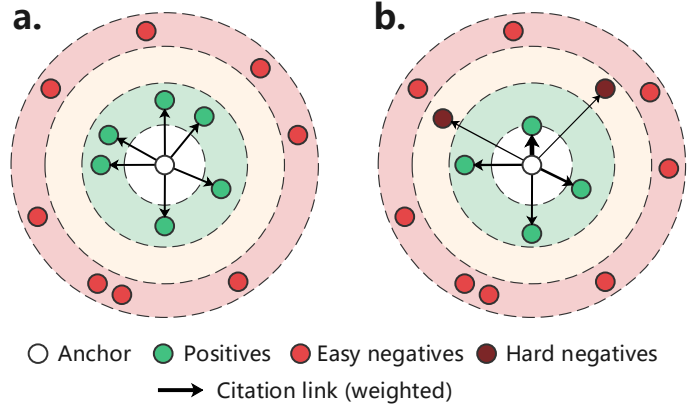

**Fig. 2.** Defining positive and negative documents using (a) raw citations vs. (b) importance-weighted citations.

The implementation of our importance-aware sampling strategy is outlined in Algorithm 1. For each anchor document, $K$ triplets are sampled to provide contrastive signals needed for model training. The hyperparameter $H$ determines the number of these $K$ triplets that contain a hard negative. This parameter plays a key role in balancing task difficulty, ensuring the model is sufficiently challenged to learn better representations without hindering convergence.

---
**Algorithm 1.** Importance-aware triplet sampling.

---
**Input:** $P$: A Dataset of scientific documents
    $N$: Total number of training samples
    $K$: Number of triplets to sample per anchor document
    $H$: Number of hard negatives to sample per anchor document
**Output:** $C$: Dataset of training samples

**Algorithm 1.** Importance-aware triplet sampling.

**Procedure:**
1: $C \leftarrow \emptyset$
2: **while** $|C| < N$ **do:**
3:    $p_a \leftarrow$ RandomSelect($P$)                // Select a random anchor document
4:    $R \leftarrow$ GetReferences($p_a$)              // Get anchor's references
5:    $R \leftarrow$ Sort($R$, key=*importance*, order=desc)    // Sort references by importance
6:    $R_{orig} \leftarrow$ Copy($R$)                  // Keep original set for sampling easy negatives
7:    $h \leftarrow 0$                           // Initialize hard negative counter
8:    **for** $i = 1$ to $K$ **do:**
9:       $p_p \leftarrow R$.pop(0)              // Select most important reference as positive
10:      **if** $h < H$ **then:**
11:         $p_n \leftarrow R$.pop(-1)          // Select least important reference as (hard) negative
12:         $h \leftarrow h + 1$
13:      **else:**
14:         $p_n \leftarrow$ RandomSelect($P \setminus R_{orig}$)   // Select non-cited document as (easy) negative
15:      **end if**
16:      $C \leftarrow C \cup \{(p_a, p_p, p_n)\}$      // Add new triplet to $C$
17:    **end for**
18: **end while**
19: remove document pairs that occur multiple times with contradictory relationships
20: **return** $C$

### 3.3 Model architecture

The triplets generated by our importance-aware sampling strategy provide essential supervisory signals for model training. This section details how these signals are integrated into a contrastive learning framework to optimize the weights of the representation learning model.

The model architecture is shown in Fig. 3. We adopt SciBERT (Beltagy et al., 2019) as the base encoder for generating document representations, given its strong performance on various scientific natural language processing tasks. For each document in a triplet ($p_a$, $p_p$, $p_n$), the title and abstract are concatenated and fed into SciBERT. The output token embeddings are then averaged using mean pooling to obtain a 768-dimensional vector representation for the document. This is a common method to aggregate token-level information into a single fixed-size vector for the entire document. Our training objective is to optimize the SciBERT using the structural supervisory signal. This optimization works by contextualizing and refining the semantic representations of scientific documents, a process guided by their importance-aware citation relationships.

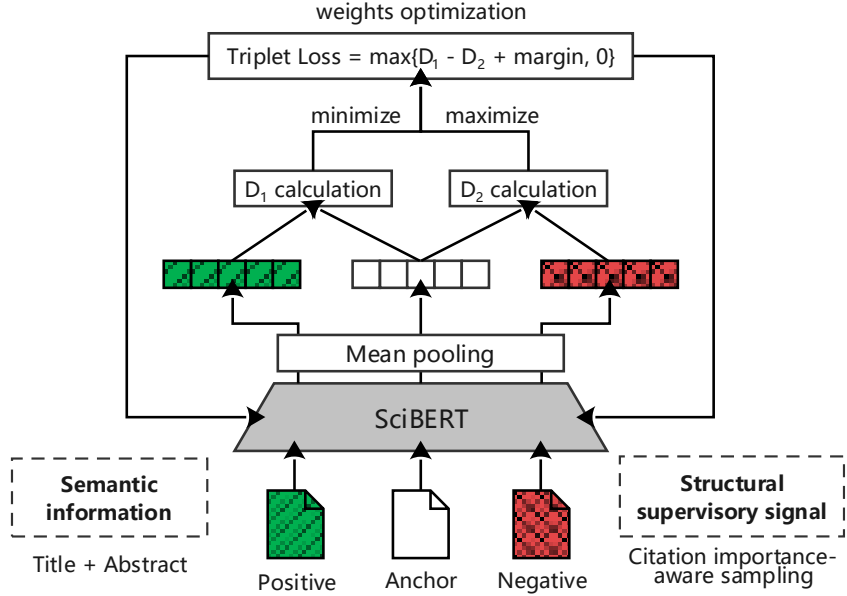

**Fig. 3.** Architecture of the citation-informed document representation learning model.

Specifically, the optimization process employs a triplet margin loss function, as shown in equation (2):

$$L(p_a, p_p, p_n) = \max\{(D(v_a, v_p) - D(v_a, v_n) + m),\ 0\} \quad (2)$$

Given a triplet of documents as input, $v_a$, $v_p$, and $v_n$ are the vector representations of the anchor, positive, and negative documents, respectively. $D$ is a distance measure of two vectors, and $m$ is the margin hyperparameter. In this study, we use the L2 norm (Euclidean distance) as our distance measure. If the positive document is already closer to the anchor than the negative document by at least the margin, the loss will be zero. Otherwise, a positive loss will drive the weight optimization process through backpropagation.

## 4 Experimental setup

### 4.1 Datasets and evaluation tasks

Several large-scale datasets are employed for model training, evaluation, and application. A brief description of these datasets is provided in Table 1.

**Table 1.** Datasets used for model training, evaluation, and application in this study.

| Purpose | Dataset | # documents | Description |
| --- | --- | --- | --- |
| Training/validation | Elsevier full-text collection | 4,339,429 | Full-text articles in Elsevier journals, collected and made accessible by CWTS. |
| Evaluation: Document representation quality | SciDocs | 226,743 | Publicly available benchmark dataset for scientific document representation. |
| Evaluation: Science mapping accuracy | PubMed | 2,941,109 | Benchmark dataset for community detection of scientific documents, made available by CWTS. |

| Purpose | Dataset | # documents | Description |
| --- | --- | --- | --- |
| Application | Web of Science | 33,230,001 | In-house version of the Web of Science database, made available by CWTS. |

*4.1.1 Training dataset*

Following our research design, full-text documents are needed for citation importance calculation. Therefore, the dataset used to train our representation model is derived from the Elsevier full-text collection, which contains 7,862,859 scientific publications published in Elsevier journals between 2000 and 2017 (Boyack et al., 2018). This dataset was collected by CWTS and made available to the first author of this paper during his visit to Leiden University. For constructing our training triplets, we focused on publications of type "Full-length Article" and their cited references, as these represent primary research outputs and generally possess an IMRaD structure. This selection results in a corpus of 4,339,429 full-length articles with a total of 67,513,443 citation relationships. From this corpus, we construct a dataset of about 240 thousand anchor documents. Following the practice in Cohan et al. (2020), we sample $K = 5$ triplets for each anchor document, resulting in a total of about 1.2 million triplets. This dataset is then split into 80% for training and 20% for validation purposes.

*4.1.2 Evaluation datasets and tasks*

The evaluation is divided into two parts, corresponding to RQ2 and RQ3 of this study. To answer RQ2, we employ the SciDocs benchmark, which is a publicly available dataset designed for evaluating the quality of scientific document representation (Cohan et al., 2020). The key idea of the SciDocs benchmark is to evaluate document representations by treating them as fixed input features for simple downstream models (e.g., linear support vector machines or shallow neural networks), or by directly using vector similarities without any further modeling. The SciDocs benchmark dataset contains 226 thousand documents and four types of evaluation tasks:

(1) **Document classification**. This task involves classifying documents into 11 MeSH top-level disease categories and 19 MAG top-level topic categories. Model performance is evaluated by the F1-score.

(2) **Citation prediction**. This task requires ranking a set of 30 candidate documents for a target document, where 5 are genuinely related through direct citations or co-citations. Model performance is evaluated by mean average precision (MAP) and normalized discounted cumulative gain (nDCG).

(3) **User activity prediction**. Similar to citation prediction, this task requires ranking a set of 30 candidate documents for a target document, where 5 are genuinely related through co-view and co-read relationships.

(4) **Document recommendation**. Based on user clickstream data from an academic search engine, this task involves ranking documents relevant to a user query. Model performance is evaluated by precision@1 (P@1) and nDCG.

To answer RQ3, we employ a benchmark dataset developed by Ahlgren et al. (2020). This dataset consists of 2,941,109 articles and reviews indexed in PubMed between 2013 and 2017. The task is to perform community detection on scientific networks constructed using different document relatedness measures. Citation based networks, where documents are connected based on citation links, and networks based on textual similarities, where each document is connected to its top 20 most similar documents, are considered. Communities in the networks are identified using the Leiden algorithm (Traag et al., 2019), a widely adopted method in science mapping studies.

The quality of the resulting community structures (referred to as clustering accuracy) is evaluated by using Medical Subject Headings (MeSH)-based document similarities as an independent evaluation criterion (Ahlgren et al., 2020; Waltman et al., 2020). Specifically, the accuracy is calculated as the average MeSH similarity between all pairs of documents within the same community, across all communities. The calculation equation is defined as follows (Waltman et al., 2020):

$$A^X = \frac{1}{N} \sum_{i,j} I(c_i^X = c_j^X) r_{ij}^{MeSH} \quad (3)$$

where $X$ denotes the document relatedness measure to be evaluated, $N$ is the total number of document pairs, $c_i^X$ and $c_j^X$ is a positive integer denoting the community to which document $i$ and $j$ belongs with respect to relatedness measure $X$. $I(c_i^X = c_j^X)$ is 1 if document $i$ and $j$ belongs to the same community, otherwise 0. $r_{ij}^{MeSH}$ is the normalized MeSH similarity of document $i$ and $j$. Furthermore, to provide a more robust and comprehensive evaluation, $A^X$ is calculated for each relatedness measure using multiple resolution parameter settings of the Leiden algorithm.

### 4.1.3 Application dataset

Finally, the model trained with our proposed importance-aware contrastive learning framework is applied to a large-scale interdisciplinary dataset consisting of all articles, reviews, and proceedings papers indexed in Web of Science between 2000 and 2022, totaling over 33 million scientific documents. The in-house version of Web of Science available at CWTS was used for this purpose and made accessible to the first author during his visit to Leiden University.

### 4.2 Baselines

For the SciDocs benchmark, eleven document embedding models are selected as baselines. Table 2 shows the comparison between these baselines and our proposed model in terms of four key features. The baselines include classical text embedding models Fasttext and Doc2vec (Bojanowski et al., 2017; Le & Mikolov, 2014), pretrained language models BERT (Devlin et al., 2019) and its scientific variant SciBERT (Beltagy et al., 2019), those further fine-tuned with contrastive learning such as SimCSE (Gao et al., 2021) and SPECTER series (Cohan et al., 2020; Singh et al., 2023), a graph representation learning model SGC (Wu et al., 2019), Microsoft's E5-base-v2 model (Wang et al., 2022) and OpenAI's text-embedding-ada-002 model (Neelakantan et al., 2022). Among these, SPECTER serves as our major baseline, as it introduced the fundamental idea of using citations as supervisory signals for contrastive learning, though without distinguishing citation importance. SPECTER2 is a successor that further employs a multi-format training strategy with adapters and a 10 times larger training dataset. In addition, we include SciNCL (Ostendorff et al., 2022) as it employs an idea similar to ours to provide more accurate supervisory signals for citation-informed contrastive learning. Training details of the above baselines are available in Appendix A.

**Table 2.** Comparison between baselines and our proposed model.

| Model | Embedding size | Citation information | Contrastive learning | Importance awareness |
|---|---|---|---|---|
| **Ours** | **768** | **Yes** | **Yes** | **Yes** |
| Fasttext | 300 | No | No | No |
| Doc2vec | 300 | No | No | No |
| BERT | 768 | No | No | No |
| SciBERT | 768 | No | No | No |
| SGC | 300 | Yes | No | No |
| SimCSE | 768 | No | Yes | No |
| SPECTER | 768 | Yes | Yes | No |
| SPECTER2 | 768 | Yes | Yes | No |
| SciNCL | 768 | Yes | Yes | No |
| E5-base-v2 | 768 | Yes | Yes | No |
| text-embedding-ada-002 | 1536 | No | Yes | No |

For the science mapping benchmark on the PubMed dataset, we compare community detection results obtained by our proposed model with results from other approaches to inter-document relationships. These baselines include traditional citation-based methods, including direct citations (DC), extended direct citations (EDC, Waltman et al., 2020), co-citations (CC), and bibliographic coupling (BC), as well as a text-based method, BM25, which computes similarity based on noun phrases extracted from titles and abstracts. Results for these baselines are taken from Ahlgren et al. (2020). We also include other PLMs in our comparison, including SciBERT (Beltagy et al., 2019) and its contrastive fine-tuned variants SPECTER (Cohan et al., 2020), SPECTER2 (Singh et al., 2023), and SciNCL (Ostendorff et al., 2022). For the BM25, SciBERT, and its variants, the scientific network is constructed by connecting each document to its 20 most similar neighbors. This number approximates the average number of references per document and represents the upper limit that our hardware can support for constructing the network on the dataset.

### 4.3 Implementation details

We implement the proposed model with PyTorch and Transformers. The model is trained on an Nvidia RTX 4090 24GB GPU. We set the learning rate to 2e-5, batch size to 8, with a gradient accumulation step equal to 4, and epochs to 2, according to the practice in Cohan et al. (2020). An AdamW optimizer and a linear scheduler are employed, which decreases the learning rate after 10% warm-up steps. The margin in the loss function and the number of hard negatives ($H$) are grid-searched based on validation performance. The final model was trained with a margin of 1 and $H$ set to 2.

Pre-trained models such as BERT, SciBERT, SPECTER, SPECTER2, SciNCL, and E5-base-v2 are obtained from the Hugging Face Model Hub. Embeddings of the text-embedding-ada-002 (Ada v2) model are generated via the OpenAI embedding API. The performance results of the remaining baselines are taken from existing studies that used the same evaluation dataset (Cohan et al., 2020; Ahlgren et al., 2020; Ding et al., 2024).

## 5 Results

To evaluate the proposed framework, this section presents the results in three parts. We first analyze the components of our citation importance measurement to understand the characteristics of the supervisory signals. Next, we evaluate the quality of the document representations generated by the model trained with our framework (RQ2). Finally, we demonstrate the effectiveness of our model in the large-scale application of science mapping (RQ3).

### *5.1 Analysis of citation importance weights*

The weights of each component in the citation importance measurement (equation 1) are estimated using the entropy weight method on the Elsevier full-text collection. These weights reflect the relative contribution of each feature to the overall citation importance score. As shown in Table 3, self-citation is the most important component (34.57%), followed by the cited frequency in Results (29.33%) and Discussion & Conclusion (24.38%) sections. By contrast, cited frequency in Introduction sections turns out to be a relatively weak indicator (11.73%) of citation importance. This suggests that citations reflecting continuity (self-citations) or related to research content are stronger indicators of citation importance, distinguishing them from the general background references typically found in the Introduction. With these weights determined, we apply them to sample training triplets and fine-tune the model.

Regarding the exclusion of citations in the Method section, our empirical findings support this choice. As will be detailed in the ablation study in Section 5.2.2, including citation frequencies in the Method section $f_M$ leads to a reduction in model performance.

**Table 3.** Estimated weights of the components of citation importance.

| Component | Mean | Std. | Entropy | Weight |
|---|---|---|---|---|
| $f_I$ – cited frequency in Introduction sections (including literature review) | 0.70 | 0.80 | 0.96 | 11.73% |
| $f_R$ – cited frequency in Results sections | 0.32 | 0.79 | 0.91 | 29.33% |
| $f_D$ – cited frequency in Discussion & Conclusion sections | 0.77 | 0.75 | 0.92 | 24.38% |
| $S$ – self-citation indicator | 0.14 | 0.34 | 0.89 | 34.57% |

### *5.2 Evaluation of document representation quality (RQ2)*

#### *5.2.1 Proposed implementation of citation importance assessment*

To answer the second research question, we evaluate the quality of the scientific document representations generated by our proposed method and several baselines using the SciDocs benchmark. The results across all four task categories are presented in Table 4.

Overall, the model trained with our proposed importance-aware sampling method achieves an average performance of 81.3, representing a noticeable improvement over the original

SPECTER model (80.0). The best overall performance is achieved by SciNCL (81.8), which employs a more sophisticated approach for constructing contrastive triplets based on document similarity. Although SciNCL also acknowledges that not all citations are equal, it operationalizes this principle differently than our proposed method. Nonetheless, our method requires fewer computational resources while remaining competitive in performance. It is important to emphasize that our primary aim is not to achieve SOTA general performance, but rather to empirically demonstrate that the scientometric principle of citation heterogeneity can be operationalized to improve the quality of scientific document representation, which in turn benefits our target application of science mapping (see Section 5.3).

By comparing the relative performance of different types of models, several key findings are observed:

(1) **PLMs that have not been optimized with contrastive learning, such as BERT and SciBERT, perform poorly.** Their average performance is even lower than that of simpler models like Fasttext and Doc2vec. Although BERT and SciBERT have been proven effective in downstream NLP tasks, their effectiveness often depends on the presence of downstream models being sophisticated enough to perform additional feature extraction and transformation on the document representations. The SciDocs benchmark, however, involves only simple models or no models at all, which places a higher demand on the intrinsic quality of the representations. As stated in previous studies (Gao et al., 2019; Li et al., 2020), the representations generated by these PLMs suffer from the anisotropy problem, limiting their expressiveness and suitability for direct similarity comparisons.

(2) **Contrastive learning significantly improves the quality of document representations.** SimCSE, which applies a simple contrastive learning strategy on the text itself, outperforms the base BERT model by a large margin (6.4 percentage points). This shows that contrastive learning can effectively mitigate the anisotropy problem by enforcing a more uniform and aligned distribution of representations in the embedding space. By training on a huge amount of data with contrastive learning, OpenAI's text-embedding-ada-002 model achieves competitive performance and even outperforms our main baseline SPECTER.

(3) **Integration of citation information into the contrastive learning framework further enhances document representation quality.** By using citation links to construct positive pairs, SPECTER substantially outperforms SimCSE, raising the average performance score from 69.8 to 80.0. Its successor, SPECTER2, which employs a multi-format training strategy with adapters and a 10 times larger training dataset, achieves an even higher performance of 81.5. This improvement is particularly evident in citation prediction and user activity prediction tasks. Similarly, Microsoft's E5-base-v2 model partly involves a citation-informed contrastive learning process. Its performance is lower than the SPECTER models, as the latter are specifically fine-tuned on scientific data.

(4) **Incorporating semantic information into the graph representation learning model is another effective strategy for enhancing document representation quality.** The SGC model employs this approach by initializing node vectors in the citation network with text representations. Its overall performance exceeds that of models relying solely on textual content. However, this approach is outperformed by text-based models that integrate citation information through contrastive learning. In addition, graph-based models are more computationally expensive and more difficult to generate representations for new documents.

(5) Finally, differentiating citations provides an additional performance boost. Our model builds upon the citation-informed contrastive learning framework of SPECTER but introduces a more nuanced supervisory signal. By requiring the model to distinguish not only between cited and non-cited documents, but also between important and perfunctory citations (i.e., hard negatives), our model achieves an overall performance of 81.3. The best overall performance is achieved by SciNCL (81.8), which first learns paper embeddings on the citation network and then samples training triplets from the "citation neighborhood". While both SciNCL and our model leverage citation heterogeneity, their implementations are fundamentally different. The key distinction lies in how the supervisory signals are derived. Our method directly ranks citation links based on a calculated importance score (see Section 3.1). In contrast, SciNCL establishes positive and negative relationships based on the similarity of papers within a pre-computed citation embedding space. As a result, all positives in the training data of our model must be cited by the anchor document, while this restriction is lifted in the more flexible "citation neighborhood" approach by SciNCL. Despite the differences in implementation, the results confirm that refining the citation-based supervisory signal is an effective way to improve the quality of scientific document representation.

**Table 4.** Model performances on the SciDocs benchmark.

| Model | Classification | | Citation prediction | | | | User activity prediction | | | | Recommendation | | Avg. |
|---|---|---|---|---|---|---|---|---|---|---|---|---|---|
| | MAG | MeSH | Cite | | Co-cite | | Co-view | | Co-read | | | | |
| | F1 | F1 | MAP | nDCG | MAP | nDCG | MAP | nDCG | MAP | nDCG | nDCG | P@1 | |
| SciNCL | 81.2 | 88.8 | 85.4 | 92.3 | 87.7 | 94.0 | 93.4 | 97.3 | 91.7 | 96.5 | 54.3 | 19.5 | 81.8 |
| SPECTER2 | 82.5 | 89.1 | 85.3 | 92.3 | 87.0 | 93.6 | 92.1 | 96.8 | 91.2 | 96.3 | 53.0 | 18.9 | 81.5 |
| Ours* | 83.3(0.1) | 89.8(0.4) | 84.5(0.2) | 92.0(0.1) | 85.4(0.4) | 92.9(0.2) | 91.1(0.2) | 96.2(0.1) | 89.5(0.2) | 95.5(0.1) | 54.9(0.2) | 20.6(0.5) | 81.3(0.1) |
| Ada v2 | 83.6 | 88.7 | 84.0 | 91.6 | 84.5 | 92.5 | 86.4 | 94.1 | 87.3 | 94.4 | 54.4 | 20.6 | 80.2 |
| SPECTER | 82.0 | 86.4 | 88.3 | 94.9 | 88.1 | 94.8 | 83.6 | 91.5 | 84.5 | 92.4 | 53.9 | 20.0 | 80.0 |
| E5-base-v2 | 82.5 | 89.0 | 80.8 | 90.1 | 81.4 | 91.1 | 80.9 | 91.5 | 84.6 | 93.4 | 53.1 | 18.1 | 78.0 |
| SGC | 76.8 | 82.7 | 91.6 | 96.2 | 84.1 | 92.5 | 77.2 | 88.0 | 75.7 | 87.5 | 52.7 | 18.2 | 76.9 |
| SimCSE | 80.4 | 75.2 | 65.6 | 82.8 | 70.2 | 85.1 | 71.2 | 85.1 | 67.8 | 83.3 | 52.6 | 18.0 | 69.8 |
| Fasttext | 70.5 | 72.3 | 70.6 | 81.5 | 72.8 | 83.9 | 69.5 | 81.6 | 73.3 | 82.4 | 51.5 | 17.3 | 68.9 |
| Doc2vec | 66.2 | 69.2 | 65.3 | 82.2 | 67.1 | 83.4 | 67.8 | 82.9 | 64.9 | 81.6 | 51.7 | 16.9 | 66.6 |
| BERT | 79.9 | 74.3 | 54.3 | 75.1 | 57.9 | 77.3 | 59.9 | 78.3 | 57.1 | 76.4 | 52.1 | 18.1 | 63.4 |
| SciBERT | 79.4 | 79.9 | 53.2 | 73.8 | 57.7 | 77.4 | 59.8 | 78.1 | 55.7 | 75.3 | 51.6 | 17.3 | 63.3 |

Note: *For our model, we report the mean performance and standard deviation (in parentheses) across 5 runs with different random seeds.

Based on the above findings, we further investigate the impact of task difficulty on model performance. In our framework, task difficulty is mainly affected by the margin in the loss function and the number of hard negatives per anchor ($H$). We perform a grid search with margin values of [0, 0.5, 1] and $H$ values ranging from 0 to 5. The results are shown in Fig. 4.

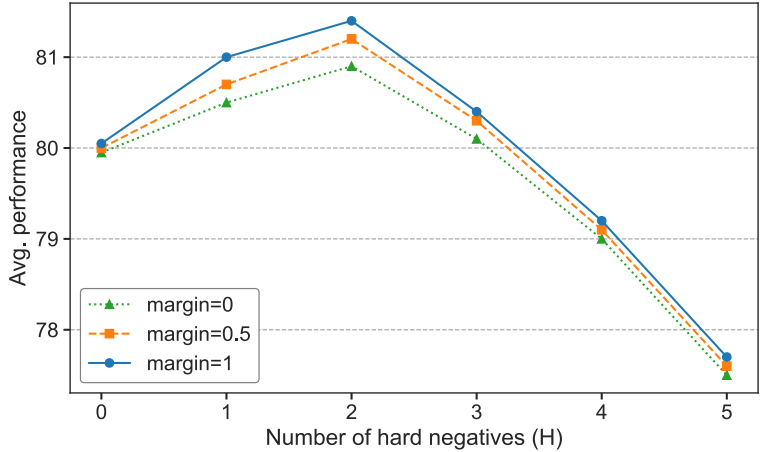

**Fig. 4.** Impact of task difficulty on model performance on the SciDocs benchmark.

An inverted U-shape relationship between task difficulty and model performance is observed in Fig. 4. The figure also reveals that the number of hard negatives ($H$) has a more substantial effect on performance than the margin. When $H$ equals 0, the task difficulty is at its lowest, as the model is only exposed to easy negatives. This results in suboptimal performance, similar to that of SPECTER. As $H$ increases to 2, the model is challenged with a moderate level of task difficulty. This requires the model to learn finer-grained distinctions between document pairs, thereby boosting the performance to a peak of 81.4. However, as $H$ increases beyond this optimal point, the task becomes too difficult and performance declines. The excessive number of hard negatives may introduce too much noise into the training signal, which negatively affects learning and harms performance. These findings suggest that the model is sensitive to task difficulty, which is an important factor to consider when designing contrastive learning frameworks. While our grid search identifies a globally optimal value for $H$ and the margin, a fixed threshold may not be ideal for all scientific fields. For instance, documents in densely cited fields might benefit from more hard negatives to learn finer-grained distinctions. Therefore, exploring adaptive strategies for selecting the number of hard negatives may be a promising direction for future studies.

### 5.2.2 Alternative implementations and ablation study

To examine the robustness of the performance improvement brought by differentiating citations, we further investigate alternative implementations of our citation importance measurement. Specifically, we focus on alternative feature combinations and weight estimation methods. Table 5 shows the SciDocs performance of models trained with data constructed by different citation importance measurements. Several insights can be gained from the results.

First, all alternative implementations of citation importance measurements improve model performance compared to our main baseline (SPECTER), which does not differentiate citations. This confirms the robustness of our citation importance-aware approach. Second, citation frequency in the Introduction section is the most important feature. Excluding it results in the largest performance drop. This is reasonable as most citations occur in the Introduction section (Bertin et al., 2016; Boyack et al., 2018). Third, there is no significant difference between the two weight estimation methods. However, in the case of the entropy weight method, including citation frequencies in the Method section leads to a performance decrease. This may be due to the high entropy weight assigned to $f_M$, while the thematic relevance of the cited papers is relatively low. In contrast, this decrease is not observed in the case of the uniform weight method. Lastly, incorporating a simple semantic feature of title similarity between citing and cited documents provides a slight performance boost under the entropy weight method. This feature may serve as an additional signal for relatedness in cases where citation features are sparse, such as references that are cited only once. Overall, these results confirm the robustness of performance improvement is confirmed.

**Table 5.** Average performance on SciDocs of models trained with data constructed by alternative citation importance measurements.

| Feature combinations | Weight estimation method | |
| --- | --- | --- |
| | Entropy weight method | Uniform weights |
| All features ($f_I, f_M, f_R, f_D, S$) | 81.20 | 81.25 |
| w/o $f_I$ | 81.09 | 81.07 |
| w/o $f_M$ | 81.31* | 81.23 |
| w/o $f_R$ | 81.20 | 81.17 |
| w/o $f_D$ | 81.14 | 81.15 |
| w/o $S$ | 81.18 | 81.21 |
| w/ title similarity[+] | 81.25 | 81.24 |

Note: *This is the proposed implementation in Section 5.2.1. [+]Title similarity is calculated as the cosine similarity between two embeddings generated by SciBERT.

## 5.3 Evaluation of science mapping accuracy (RQ3)

To address the third research question, we evaluate the accuracy of community detection, an important task in science mapping, based on document relationships established by different methods. This evaluation uses the PubMed benchmark developed by Ahlgren et al. (2020). A network of scientific documents is first constructed, with links between documents established based on citations or similarity. Communities are then identified using the Leiden algorithm. The quality of these communities, referred to as clustering accuracy, is measured by the average MeSH similarity between all pairs of documents belonging to the same community. The similarity calculation process can be found in Ahlgren et al. (2020). Since the MeSH descriptors of a paper are annotated by domain experts, they can be seen as an expert-validated ground truth for assessing the topic coherence of the algorithmically constructed communities. The results at different levels of granularity are presented in Fig. 5.

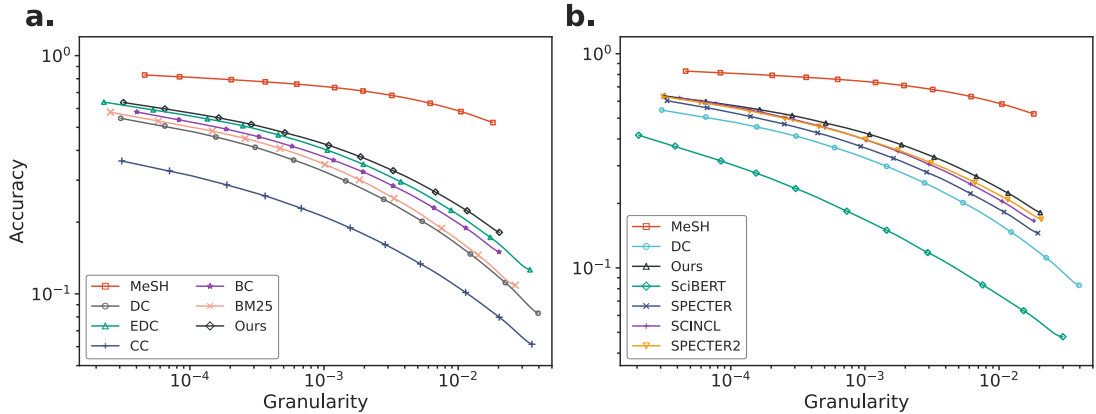

**Fig. 5.** Community detection performance using (a) traditional methods and (b) other representation learning models. The line annotated as "MeSH" represents the theoretical upper bound of clustering accuracy.

As shown in Fig. 5, our proposed model consistently outperforms both traditional and other representation learning methods in the community detection task across all granularity levels. As granularity increases, the performance gap between our model and baselines enlarges. This indicates the strength of our model in fine-grained community detection. A consistent downward trend is also observed, which is an expected outcome according to the definition of accuracy (see equation 3). It is worth noting that the line annotated as "MeSH" represents the theoretical upper bound of the accuracy (Ahlgren et al., 2020). This line serves only as a reference and is not one of the methods being evaluated.

In Fig. 5a, all methods outperform direct citations (DC) except co-citations (CC), which highlights the limitation of raw citation networks in science mapping. However, Fig. 5b shows that SciBERT representations perform even worse than the DC baseline. This confirms that representations from PLMs without contrastive fine-tuning are of lower quality. The inter-document relationships established based on representation similarity are less accurate. In contrast, SPECTER, which incorporates citation information through contrastive learning, surpasses the DC baseline. Our proposed model further improves upon SPECTER by using a more refined, importance-aware supervisory signal, achieving the highest accuracy. Interestingly, although SPECTER2 and SciNCL perform better on the SciDocs benchmark, they are slightly outperformed by our model in this community detection evaluation for science mapping.

To provide a clear overview of the overall performance, Fig. 6 summarizes the relative accuracy of each method with a single score. This metric is calculated by comparing the accuracy of each method against the average accuracy of all methods at all granularity levels. The final score is the average of these relative comparisons, where a value of 1 represents the average performance level (Ahlgren et al., 2020).

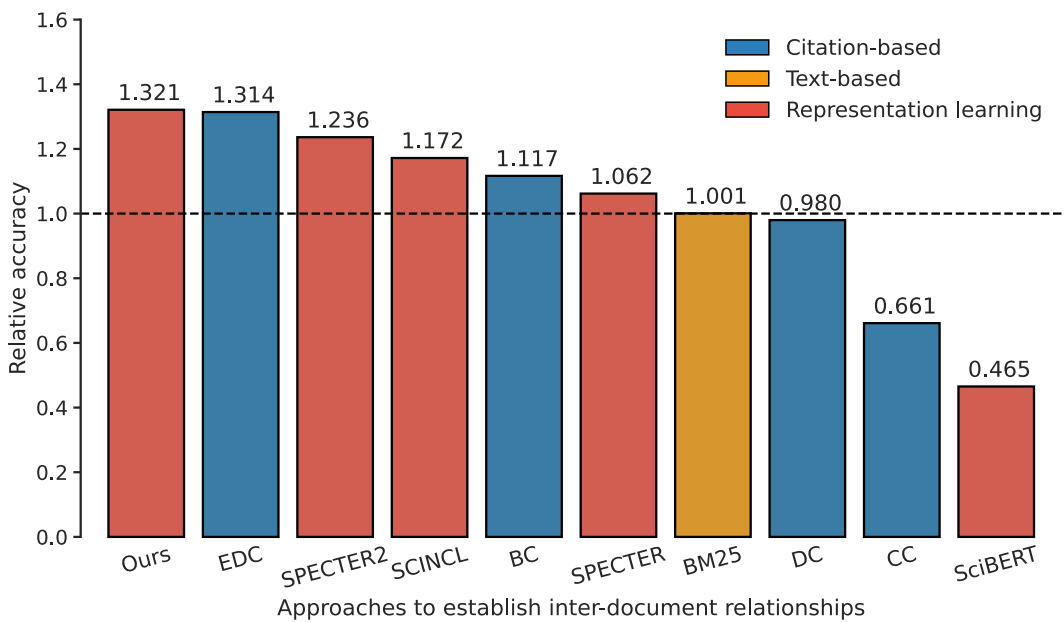

**Fig. 6.** Relative performance of all methods and models on the PubMed benchmark.

Fig. 6 shows that our proposed model achieves the highest relative accuracy of 1.321, followed by citation-based method EDC (1.314), and the representation learning models SPECTER2 (1.236) and SciNCL (1.172). In contrast, CC and SciBERT perform well below the average, with scores of 0.661 and 0.465, respectively. These results indicate that integrating semantic information and citation-based supervisory signals is essential for establishing accurate inter-document relationships. Moreover, refining the supervisory signal by differentiating citation links further improves performance.

## 5.4 Large-scale application: Mapping the global structure of science (RQ3)

Having demonstrated the superior performance of our proposed model on benchmark tasks, we further apply it to a large-scale, real-world scenario. The goal is to create a comprehensive map of science that facilitates the understanding of the global scientific landscape. To this end, we first generate representations for over 33 million scientific documents indexed in the Web of Science database from 2000 to 2022, covering all disciplines. Next, we construct a network of scientific documents by connecting each document to its 20 most similar neighbors, based on the cosine similarity of their vector representations. The documents are then clustered into communities that reflect the structure of science at the granularity of research topics (Sjögårde & Ahlgren, 2018). Finally, a global map of science is obtained by dimensional reduction and visualization of these research topics.

### 5.4.1 The underlying scientific document network

Before visualizing the global map of science, we first analyze the structural characteristics of the underlying network. Table 6 compares our similarity-based network with a direct citation network and a random network with the same density. It is clear that our network is a single connected component, while the direct citation network is highly fragmented, breaking into 903,780 separate components and leaving more than 880,000 documents as isolated nodes.

**Table 6.** Structural characteristics of the scientific document networks constructed using different methods.

| Network property | Our network | Citation network | Random network |
| --- | --- | --- | --- |
| Nodes | 33,230,001 | 33,230,001 | 33,230,001 |
| Edges | 515,352,109 | 674,919,425 | 515,344,156 |
| Average degree | 31.02 | 40.62 | 31.02 |
| Density | 9.33e-7 | 1.22e-6 | 9.33e-7 |
| Connected | True | False | True |
| Connected components | 1 | 903,780 | 1 |
| Isolated nodes | 0 | 880,913 | 0 |
| Clustering coefficient | 0.16 | 0.02 | 9.43e-7 |
| Avg. shortest path length* | 7.91 | 5.39 | 5.37 |

Note: *Due to the large scale of the network, the average shortest path length is estimated by sampling (N=10,000).

Our network also exhibits a much stronger community structure. Compared to both the citation network and the random network, our network has a significantly higher clustering coefficient. Interestingly, the average shortest path length of our network is also higher. This combination suggests the presence of local community structures. In other words, the network contains densely connected communities, but the connections between these communities are relatively sparse. Such characteristics are ideal for community detection and form an important foundation for accurate science mapping.

To reveal more detailed differences between our network and the traditional citation network, we conduct a comparative analysis of network edges. As illustrated in Fig.7a, about 13% of edges are shared between our network and the citation network, indicating a meaningful degree of agreement. By integrating semantic and citation information, our model does not simply replicate the citation network, but rather refines it by ignoring perfunctory citation links and connecting thematically related documents. Fig. 7b shows the comparison of our network and the random network. The negligible overlap of only 446 edges confirms that the agreement with the citation network is not caused by random network structure.

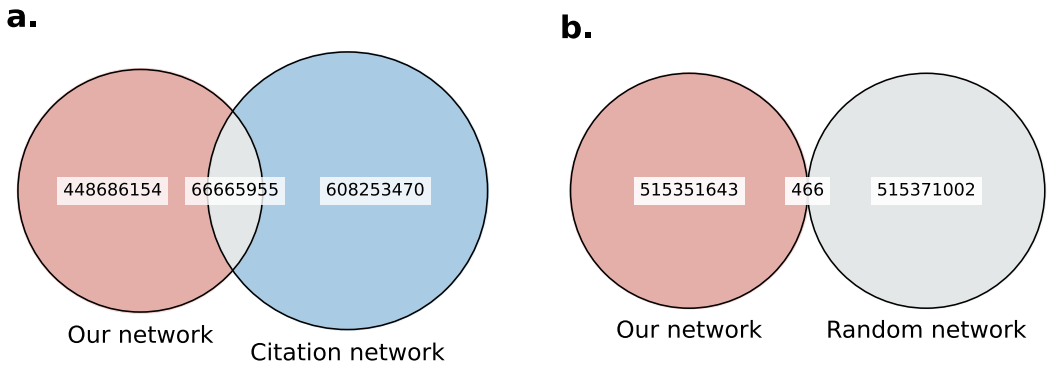

**Fig. 7.** Edge overlap between our network and (a) the citation network, and (b) the random network.

To illustrate the model's ability to identify thematically related literature, we present two case studies using papers published in *Information Processing & Management*. Table 7 lists the 10 most similar papers to the focal paper titled "*A probabilistic framework for integrating* *sentence-level semantics via BERT into pseudo-relevance feedback*" (Pan et al., 2022). Among them, 4 papers are cited by the focal paper. The other papers, although not cited, seem strongly thematically related, demonstrating our model's ability to accurately discover relevant literature beyond citation links. In contrast, Table 8 shows the 10 most similar papers to the focal paper titled "*Chatbot as an emergency exist: Mediated empathy for resilience via human-AI interaction during the COVID-19 pandemic*" (Jiang et al., 2022). None of these papers are cited by the focal paper. This discrepancy is likely due to the novelty of the second paper's research topic. For emerging topics like the use of AI technologies during the pandemic, it is challenging to keep track of the latest relevant literature, especially those published by researchers from different disciplines. In such cases, our model is particularly effective in identifying coherent research fronts where traditional citation links are absent. It is worth noting that the similarity search is conducted across the entire Web of Science corpus of about 33 million documents, indicating the effectiveness and scalability of our model.

**Table 7.** The 10 most similar papers to the focal paper "*A probabilistic framework for integrating sentence-level semantics via BERT into pseudo-relevance feedback*"

| Title | Similarity | Is cited |
| --- | --- | --- |
| A Pseudo-relevance feedback framework combining relevance matching and semantic matching for information retrieval | 0.928 | Yes |
| Revisiting Rocchio's Relevance Feedback Algorithm for Probabilistic Models | 0.920 | Yes |
| Pseudo-relevance feedback query based on Wikipedia | 0.909 | No |
| The effect of low-level image features on pseudo relevance feedback | 0.908 | No |
| TopPRF: A Probabilistic Framework for Integrating Topic Space into Pseudo Relevance Feedback | 0.908 | Yes |
| Proximity-based Rocchio's Model for Pseudo Relevance Feedback | 0.908 | Yes |
| Promoting Divergent Terms in the Estimation of Relevance Models | 0.904 | No |
| Pseudo-Relevance Feedback Based on Locally-Built Co-occurrence Graphs | 0.904 | No |
| A learning to rank approach for quality-aware pseudo-relevance feedback | 0.904 | No |
| Word-embedding-based pseudo-relevance feedback for Arabic information retrieval | 0.903 | No |

**Table 8.** The 10 most similar papers to the focal paper "*Chatbot as an emergency exist: Mediated empathy for resilience via human-AI interaction during the COVID-19 pandemic*"

| Title | Similarity | Is cited |
| --- | --- | --- |
| It Is Me, Chatbot: Working to Address the COVID-19 Outbreak-Related Mental Health Issues in China. User Experience, Satisfaction, and Influencing Factors | 0.888 | No |
| AI Chatbot Design during an Epidemic like the Novel Coronavirus | 0.886 | No |
| User Experiences of Social Support From Companion Chatbots in Everyday Contexts: Thematic Analysis | 0.884 | No |

| Title | Similarity | Is cited |
|---|---|---|
| Qualitative exploration of digital chatbot use in medical education: A pilot study | 0.883 | No |
| Investigating differential effects of socio-emotional and mindfulness-based online interventions on mental health, resilience and social capacities during the COVID-19 pandemic: The study protocol | 0.881 | No |
| Employing a Chatbot for News Dissemination during Crisis: Design, Implementation and Evaluation | 0.881 | No |
| Perceptions of Cognitive and Affective Empathetic Statements by Socially Assistive Robots | 0.880 | No |
| The Role of AI Chatbots in Mental Health Related Public Services in a (Post)Pandemic World: A Review and Future Research Agenda | 0.880 | No |
| Empathy through the Pandemic: Changes of Different Emphatic Dimensions during the COVID-19 Outbreak | 0.879 | No |
| Empathy Not Quarantined: Social Support via Social Media Helps Maintain Empathy During the COVID-19 Pandemic | 0.879 | No |

*5.4.2 Visualization of the global map of science*

We apply the Leiden algorithm to identify communities at the granularity of research topics. The value of resolution is determined by the method proposed by Sjögårde & Ahlgren (2018). This results in 101,022 research topics for our dataset, consistent with previous research on mapping global science (Klavans & Boyack, 2017). For each research topic, a vector representation is obtained by averaging all document representations of the corresponding community.

To create the global map of science, we project the 768-dimensional topic representations into a two-dimensional space using the UMAP algorithm (McInnes et al., 2018). UMAP is a non-linear dimensionality reduction technique that is particularly effective at preserving both the local and global structure of data. The global map of science is visualized in Figures 8 to 10[*], each with different overlay colors. Each dot on the map represents a research topic. The spatial arrangement of dots in the map reveals the global intellectual structure of science, with related research topics positioned closely together.

Fig. 8 shows the map colored by five broad fields of science, derived from the journal classification of the documents within each research topic. For topics containing documents from multiple fields, the color is determined by the most frequent field. This visualization confirms the usability of our model for science mapping, as research topics belonging to the same field are clustered together. The alignment with traditional disciplinary boundaries indicates that the inter-document relationships captured by our model are meaningful. At the same time, the boundaries between different colored regions, especially near the center of the map, represent the interfaces between disciplines where interdisciplinary research is likely to take place.

---

[*] High-resolution versions of these figures are available at https://doi.org/10.5281/zenodo.16950694.

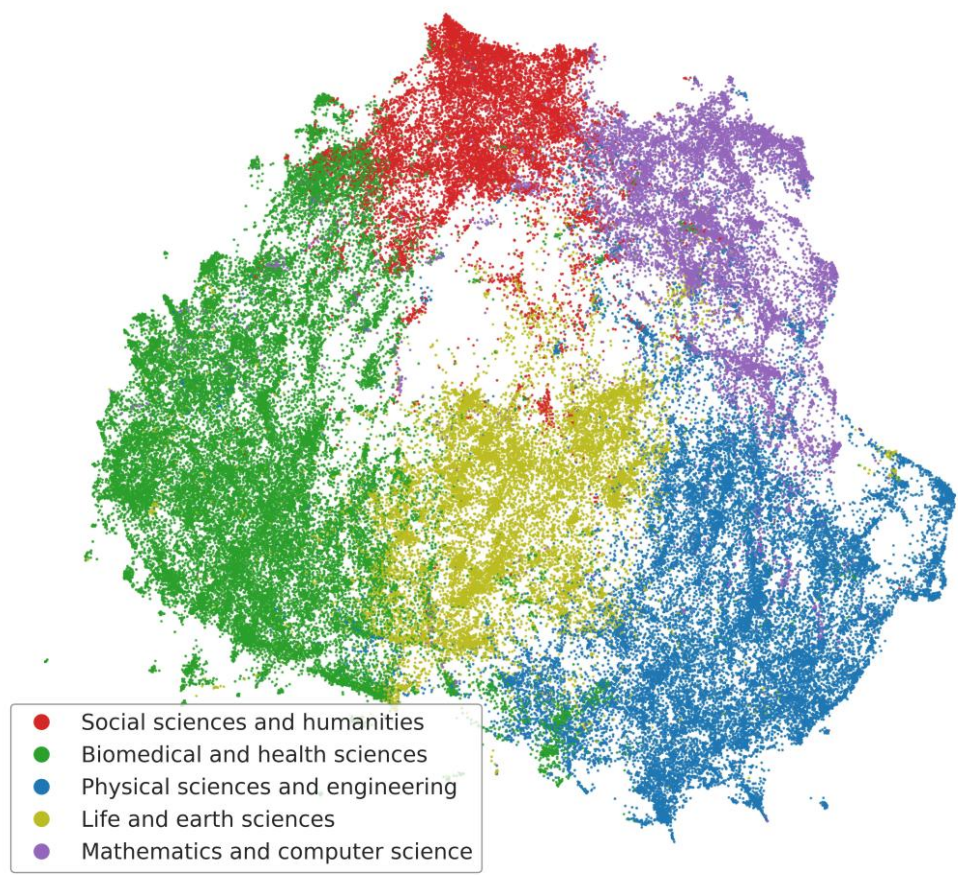

**Fig. 8.** Global map of science at the granularity of research topic, colored by the main fields of the research topics.

This hypothesis is further supported by Fig. 9, which overlays the map with the interdisciplinarity of each research topic. The Web of Science subject categories are employed as the discipline classification system. Interdisciplinarity is calculated as the product of diversity, balance, and disparity of the subject categories of the documents within each research topic (Leydesdorff et al., 2019). It is evident that interdisciplinary topics are concentrated at the boundaries between broad fields. Research topics located in peripheral areas of the map are generally less interdisciplinary than those near the center of the map. However, highly interdisciplinary topics also exist within individual fields, particularly in the social sciences and humanities.

Finally, Fig. 10 provides a temporal view of the map, where each research topic is colored by the average publication year of its associated documents. This overlay captures the evolution of science and the emergence of recent research topics. A prominent example is the dense, dark red group of research topics at the center of the map. These research topics are related to COVID-19 research, which emerged rapidly around the year 2020. By looking at these research topics in Fig. 9 and Fig. 8, we can see that COVID-19-related studies are at the intersection of three broad fields of science, including biomedical and health sciences, life and earth sciences, and social sciences and humanities. This case study demonstrates that our map captures both the disciplinary context and collaborative nature of a recent and rapid response by the scientific community to a public emergency.

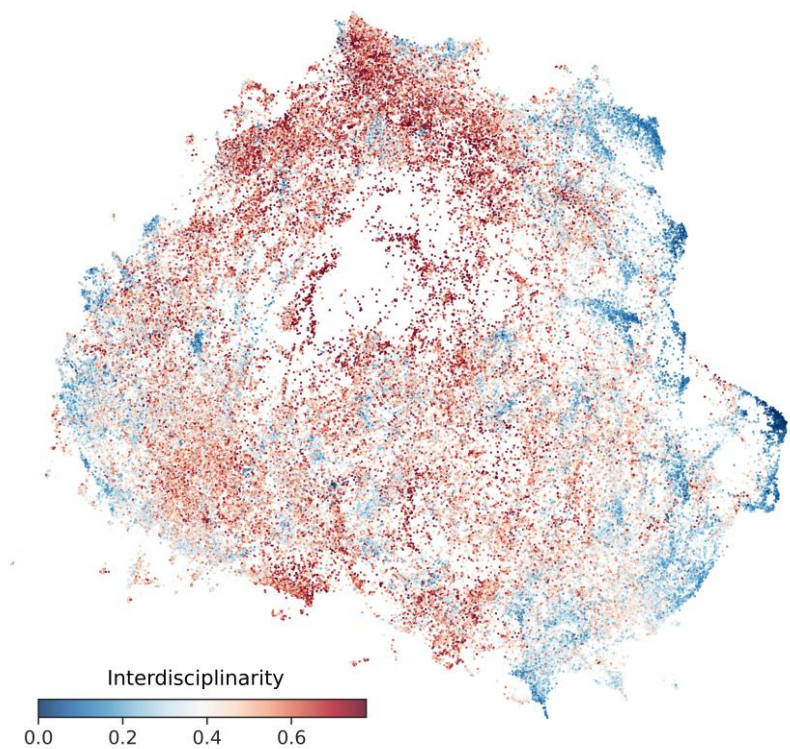

**Fig. 9.** Global map of science at the granularity of research topic, colored by the interdisciplinarity of the research topics.

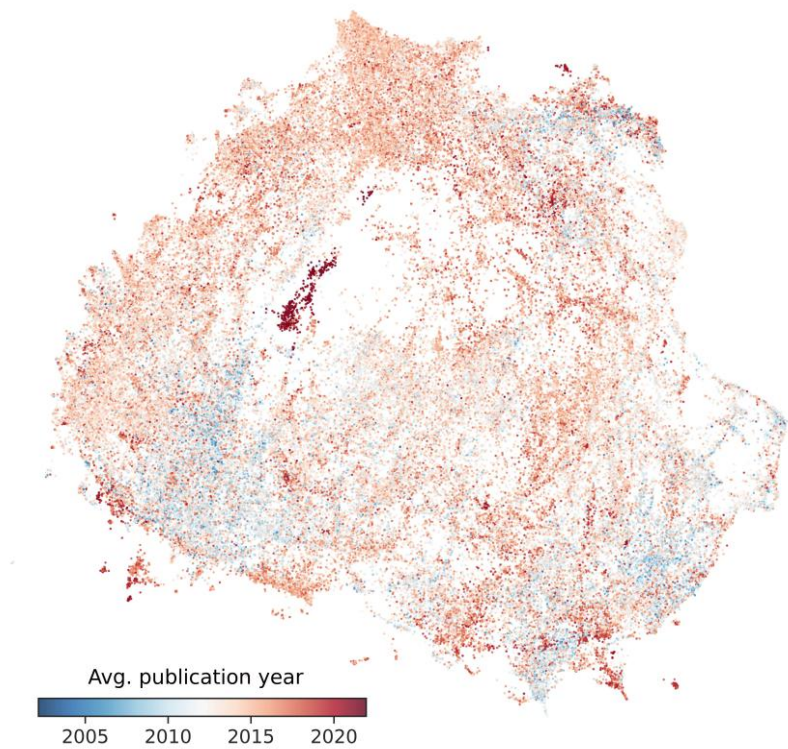

**Fig. 10.** Global map of science at the granularity of research topic, colored by the average publication year of the research topics.

# 6 Discussion

## *6.1 Implications*

From a theoretical perspective, our work offers several key insights for scientific document representation. First, we provide strong empirical validation for the principle that not all citations are equal (Zhu et al., 2015). The heterogeneity of citations can be leveraged to enhance the quality of scientific documentation representation. Second, by proposing an importance-aware sampling strategy, we demonstrate how the differences in citation importance can be operationalized into supervisory signals. With this sampling strategy, the original citation-informed contrastive learning framework can be improved. Third, our study reveals an inverted U-shaped relationship between task difficulty and document representation quality. While introducing hard negatives initially improves performance, the model is unable to learn if the tasks become too difficult. Therefore, carefully tuning task difficulty is an essential step in developing contrastive learning frameworks.

From a computational perspective, our framework offers distinct advantages in efficiency, especially when compared to graph-based alternatives. While SOTA models like SciNCL also leverage citation heterogeneity, they rely on training graph representation learning models over the entire citation network to obtain initial embeddings. This process is computationally intensive and challenging to scale to tens of millions of documents. In contrast, our approach is designed to be lightweight and scalable. Computing the metadata-based citation importance scores takes less than two minutes using only one CPU core. Although the sampling process is more time-consuming, it can be easily parallelized across multiple CPU cores. Model training is more demanding but can be completed on a single high-end consumer GPU (Nvidia RTX 4090 24GB) within 10 hours per epoch. This makes it feasible for researchers to develop and evaluate alternative approaches to citation differentiation on a reasonably powerful personal computer. While large-scale applications such as computing pairwise document similarities are computationally intensive, this is a one-time cost. The resulting representations and science maps can be stored and reused in future studies.

Furthermore, our framework improves the model performance with minimal sacrifice to interpretability. Unlike graph neural networks that often produce opaque embeddings, our approach relies on explicit scientometric features to guide the learning process. By defining hard negatives as perfunctory citations with low importance scores, we provide a transparent rationale for the model's decision to discount specific relationships. This clear definition helps researchers to understand what the model is learning, offering a certain level of transparency that is often missing in complex deep learning models.

Practically, the value of this study is most clearly demonstrated in its application-oriented context. Our central objective is to improve the accuracy of science mapping, which has been achieved in this study through generating higher-quality document representations and establishing more accurate inter-document relationships. For individual researchers, these relationships can enhance literature discovery by accurately identifying thematically related papers regardless of citations, directly supporting more relevant research recommendation systems. This is also valuable for navigating emerging or interdisciplinary research topics where citations are sparse. For policymakers and research managers, the improved representations enable the creation of more accurate science maps. These maps can offer a holistic view of science, which helps to understand disciplinary structures, identify research fronts, perform trend analysis, and track the evolution of research topics over time.

## *6.2 Limitations and future work*

This study has several limitations that suggest directions for future research.

First, the proposed measurement of citation importance is based on structural and metadata features. While this measurement is effective and scalable, it does not capture the semantic feature of a citation link. Our experiment with title similarity shows that even simple semantic signals can be beneficial (Section 5.2.2). More complex features in the citation context, however, were computationally impractical to include in this study. Future work could develop lightweight classification models to extract information about the function or intent of a citation link. The embeddings of citation sentences may also provide helpful information to differentiate citations. This information could then be integrated as an additional feature in our importance score calculation, which further refines the supervisory signal.

Second, our framework relies solely on direct citations as supervisory signals of document relatedness. Other inter-document relationships, such as co-citation and bibliographic coupling, may also be valuable for contrasting learning. Additionally, this study focuses specifically on the thematic relationship between documents. For this reason, citations from the Method section were excluded. However, it would also be interesting to construct supervisory signals based solely on methodological relatedness. A model fine-tuned with this signal might generate method-specific document representations, which can be useful for identifying different application scenarios of a research method.

Third, since our model was trained on proprietary Elsevier full-text articles, its transferability to other data sources merits further investigation. While our evaluations have shown competitive performance on benchmarks from different sources (S2ORC and PubMed), performance may decrease when the model is applied to corpora with a wider or different coverage, such as preprint servers (e.g., arXiv). A comprehensive investigation into such cross-corpus transferability remains an important direction for future work.

Finally, in the large-scale science mapping application of our model, the scientific network is constructed by connecting each document to a fixed number of its most similar neighbors (n = 20). This threshold is chosen based on the average number of references and hardware limitations. However, using a fixed value may omit relevant connections for documents in dense thematic areas, while introducing noise for documents in sparse regions. Future work is encouraged to explore an adaptive thresholding method for network construction, taking into account the heterogeneity of different disciplines and research areas.

## 7 Conclusion

In this study, we propose a citation importance-aware contrastive learning framework to enhance scientific document representation for large-scale science mapping. This framework is inspired by a fundamental scientometric principle that not all citations are equal. By developing a citation importance measurement and an importance-aware sampling strategy, we transform raw citations into more nuanced supervisory signals. Our approach introduces hard negatives into the contrastive learning process, defined as documents that are cited by the focal document but have low importance. The effectiveness of our framework is validated on two public datasets: SciDocs, for evaluating document representation quality, and PubMed, for evaluating science mapping accuracy. Results show that our model achieves improved performance on both benchmark datasets. To further demonstrate the validity and scalability of our approach, we apply the trained model to over 33 million documents from the Web of Science database. The resulting science map accurately visualizes the global and local intellectual structures of science and highlights interdisciplinary areas. Future work could focus on alternative approaches to differentiate citations, augmenting our measurement with semantic features, and exploring other types of inter-document relationships for contrastive learning. It would also be valuable to further investigate the practical validity and usefulness of our constructed map of science, particularly its local structures, in real-world bibliometric applications such as topic delineation and field-specific mapping.

## CRediT authorship contribution statement

**Zhentao Liang:** Conceptualization, Methodology, Data curation, Formal analysis, Investigation, Writing – original draft. **Nees Jan van Eck:** Methodology, Data curation, Investigation, Writing – review & editing. **Xuehua Wu:** Conceptualization, Investigation, Writing – review & editing. **Jin Mao:** Conceptualization, Methodology, Writing – review & editing. **Gang Li:** Conceptualization, Investigation, Writing – review & editing.

## Acknowledgements

This study was funded by the National Natural Science Foundation of China (Grant Nos. 72504210, 72174154, 71921002) and the open research fund of the Data and Knowledge Science Laboratory, National Science Library (Chengdu), Chinese Academy of Sciences (Grant No. CLASLKJ202502).

## Data availability

The data used in this study are proprietary and were accessed under a license agreement with Elsevier and Clarivate. The authors are not allowed to share the data.

## Appendix A. Training details of the baseline models

The authors did not train the baseline models used in this paper. To clarify potential sources of performance variation, we provide details on the scope and size of the training data used for each model.

(1) **Doc2Vec:** We adopted the performance of Doc2Vec on SciDocs from Cohan et al. (2020). Doc2Vec was trained on a subset of the Semantic Scholar corpus used to train SPECTER.

(2) **Fasttext:** We adopted the performance of Fasttext on SciDocs from Cohan et al. (2020). Fasttext was trained on a corpus of about 3.1 billion tokens from scientific papers.

(3) **SGC:** We adopted the performance SGC on SciDocs from Cohan et al. (2020). SGC was trained on a citation network derived from scientific papers in the Semantic Scholar corpus.

(4) **BERT:** We obtained the pre-trained version of BERT from the Hugging Face Model Hub. BERT was trained on BookCorpus, a dataset consisting of 11,038 unpublished books and English Wikipedia. The training data of BERT contains about 3,300 million words (Devlin et al., 2019).

(5) **SciBERT:** We obtained the pre-trained version of SciBERT from the Hugging Face Model Hub. SciBERT was trained on papers in the Semantic Scholar corpus, consisting of about 1.14 million papers with 3.1 billion tokens (Beltagy et al., 2019).

(6) **SimCSE:** We adopted the performance of SimCSE on SciDocs from Ding et al. (2024). SimCSE was trained on 1 million sentences from English Wikipedia (Gao et al., 2021).

(7) **SPECTER:** We obtained the pre-trained version of SPECTER from the Hugging Face Model Hub. SPECTER was trained on about 684 thousand training triplets of scientific papers in the Semantic Scholar corpus (Cohan et al., 2020).

(8) **SPECTER2:** We obtained the pre-trained version of SPECTER2 from the Hugging Face Model Hub. SPECTER2 was trained on about 6.2 million training triplets of scientific papers in the Semantic Scholar corpus (Singh et al., 2023).

(9) **SciNCL:** We obtained the pre-trained version of SciNCL from the Hugging Face Model Hub. SciNCL was trained on 684 thousand training triplets of scientific papers in the Semantic Scholar corpus (Ostendorff et al., 2022).

(10) **E5-base-v2:** We obtained the pre-trained version of E5-base-v2 from the Hugging Face Model Hub. E5 was trained on 270 million text pairs derived from multiple data sources (Wang et al., 2022), including Reddit, Wikipedia, Semantic Scholar corpus, etc.

(11) **Ada-v2:** We obtained the Ada-v2 embeddings of SciDocs papers using the OpenAI API service. Details of the scope and size of the training data are not provided by OpenAI.